\documentclass[12pt]{article}
    \newcommand{\be}[1]{\begin{equation}\label{#1}}
    \newcommand{\ba}[1]{\begin{eqnarray}\label{#1}}
    \newcommand{\ep}[1]{\epsilon_{#1}}
    \newcommand{\de}[1]{\delta_{#1}}

    \newcommand{\pa}[1]{\left(#1\right)}
    \newcommand{\paq}[1]{\left[#1\right]}
    
    \newcommand{\av}[1]{\langle#1\rangle}
    \newcommand{\M}{{\rm M_{\rm P}}}
     \newcommand{\z}{{\zeta}}
    \def\ee{\end{equation}}
    \def\ea{\end{eqnarray}}

\usepackage{graphicx,latexsym}
\usepackage{graphicx,epsf}
\usepackage{color}
\usepackage{epsfig}
\usepackage{amsmath}
\usepackage[T1]{fontenc}
\usepackage[utf8]{inputenc}
\usepackage{authblk}
\begin{document}
\title{Signatures of Quantum Gravity in a Born-Oppenheimer Context}

\author[1,2]{Alexander Y. Kamenshchik\thanks{Alexander.Kamenshchik@bo.infn.it}}
\author[1]{Alessandro Tronconi\thanks{Alessandro.Tronconi@bo.infn.it}}
\author[1]{Giovanni Venturi\thanks{Giovanni.Venturi@bo.infn.it}}
\affil[1]{Dipartimento di Fisica e Astronomia and INFN, Via Irnerio 46,40126 Bologna,
Italy}
\affil[2]{L.D. Landau Institute for Theoretical Physics of the Russian
Academy of Sciences, Kosygin str. 2, 119334 Moscow, Russia}
\date{}
\renewcommand\Authands{ and }
\maketitle

\begin{abstract}
We solve a general equation describing the lowest order corrections arising from quantum gravitational effects to the spectrum of cosmological fluctuations. The spectra of scalar and tensor perturbations are calculated to first order in the slow roll approximation and the results are compared with the most recent observations. The slow roll approximation gives qualitatively new quantum gravitational effects with respect to the pure de Sitter case.
\end{abstract}

\section{Introduction}
The latest Planck mission results \cite{cmb} provide the most accurate constraints available till now to inflationary dynamics \cite{inflation}. So far the slow roll (SR) mechanism has been confirmed to be a paradigm capable of reproducing the observed spectrum of cosmological fluctuations and the correct tensor to scalar ratio \cite{Stewart:1993bc}. In spite of the increased precision of observations no evident signals for quantum gravity can be extracted from the Planck data. The Inflationary period is the cosmological era describing the transition from the quantum gravitational scale down to the hot big bang scale and should, somewhere, exhibit related  peculiar features. During such a transition the cosmological perturbations with the longest wavelength are expected to be affected more by quantum gravitational effects since they exit the horizon at the early stages of inflation and are exposed to high energy and curvature effects for a longer period of time. Quite interestingly a loss of power with respect to the expected flatness for the spectrum of cosmological perturbations can be extrapolated from the data at large scales. Unfortunately such a feature (evident already in the WMAP results) exhibits large errors due to cosmic variance and, till now, its relevance seems to have been overlooked. \\
In this paper we estimate the effects of quantum gravity using the Wheeler-DeWitt equation \cite{DeWitt}. We calculate, for a realistic inflationary model, the spectrum of scalar and tensor perturbations to the first order in the SR approximation. Our approach is formally analogous to that introduced in a previous paper \cite{K} where the quantum effects on scalar perturbations evolving on a de Sitter background were estimated (similar results for the de Sitter background were also obtained in \cite{Kiefer} using a different approach). Finally the results are compared with observations. Let us emphasize than we consider a canonical quantization of Einstein gravity leading to the WDW equation, this is what we mean by quantum gravity. This is quite distinct to the introduction of so-called trans-Planckian effects through ad hoc modifications of the dispersion relation \cite{Martin} and/or the initial conditions \cite{inicond}.\\
The article is organized as follows: in section 2 we review the main equations describing the dynamics of cosmological perturbations and introduce the master equation governing the dynamics of such perturbations in the presence of quantum gravitational effects. In section 3 we introduce the slow-roll (SR) formalism. In section 4 we evaluate the quantum gravitational corrections to the master equation for scalar perturbations and obtain a general approximate solution to this equation. Subsequently some particular solutions associated with different initial conditions (vacuum choices) are discussed. In section 5 the case of tensor perturbations is addressed. In section 6 our general results are compared with observations and the effects of the quantum gravitational corrections are estimated. Finally in section 7 we illustrate our conclusions.
\section{Basic equations}
The inflaton-gravity system is described by the following action
\be{fullaction}
S=\int d\eta d^3x\sqrt{-g}\paq{\frac{\M^2}{2}R-\frac{1}{2}\partial_\mu\phi\partial^\mu\phi-V(\phi)}
\ee
which can be decomposed into a homogeneous part plus fluctuations around it. The fluctuations of the metric $\delta g_{\mu\nu}(\vec x,\eta)$ are defined by
\be{metricsc}
g_{\mu\nu}=g_{\mu\nu}^{(0)}+\delta g_{\mu\nu}
\ee  
where $g_{\mu\nu}^{(0)}=\rm{diag}\paq{a(\eta)^2\pa{1,-1,-1,-1}}$. Only the scalar and the tensor fluctuations ``survive'' the inflationary expansion: $\delta g=\delta g^{(S)}+\delta g^{(T)}$. The scalar fluctuations of the metric can be defined as follows
\be{metricsc2}
\delta g_{\mu\nu}=a(\eta)^2\left(
\begin{matrix} 
2 A(\vec x,\eta) & -\partial_i B(\vec x,\eta)\\
\\
-\partial_i B(\vec x,\eta) & 2\delta_{ij}\psi(\vec x,\eta)-D_{ij}E(\vec x,\eta)
\end{matrix}\right)
\ee
with $D_{ij}\equiv \partial_i\partial_j-\frac{1}{3}\delta_{ij}\nabla^2$. These four degrees of freedom (d.o.f.) mix with the inflation fluctuation $\delta \phi(\vec x,\eta)$ defined by $\phi(\vec x,\eta)\equiv \phi_0 (\eta)+\delta \phi(\vec x,\eta)$. Finally the scalar sector can be collectively described by a single field $v(\vec x,\eta)$ which, in the uniform curvature gauge, is given by $v(\vec x,\eta)=a(\eta)\delta \phi(\vec x,\eta)$. Its Fourier transform, $v_k$ can then be decomposed into two parts: $v_{1,k}\equiv \rm{Re}\pa{v_k}$ and $v_{2,k}\equiv \rm{Im}\pa{v_k}$. \\ 
The tensor fluctuations are gauge invariant perturbations of the metric and are defined by
\be{metrich}
ds^2=a(\eta)^2\paq{d\eta^2-\pa{\delta_{ij}+h_{ij}}dx^idx^j}
\ee
with $\partial^i h_{ij}=\delta^{ij}h_{ij}=0$. For each direction of propagation of the perturbation $k^i$
the above conditions on $h_{ij}$ with the requirement $g_{\mu\nu}=g_{\nu\mu}$ give seven independent equations for the components of the tensor perturbations leading to only two remaining polarization physical degrees of freedom $h^{(+)}$ and $h^{(\times)}$. Then, on defining $v_{1,k}^{(\lambda)}\equiv \frac{a \M}{\sqrt{2}}\rm{Re}\pa{h_k}$ and $v_{2,k}^{(\lambda)}\equiv \frac{a \M}{\sqrt{2}}\rm{Im}\pa{h_k}$ one can describe the tensor perturbations in a manner similar to the scalar perturbations.\\
In what follows we shall illustrate in detail a point which is often glossed over: namely the fact that on working in a flat 3-space and considering both homogeneous and inhomogeneous quantities one must introduce an unspecified length $L$. Indeed the effective action of the homogeneous inflaton-gravity system plus the inhomogeneous perturbations finally is \cite{MukMald}
\ba{act}
S&=&\int d\eta\left\{L^3\paq{-\frac{\M^2}{2}a'^2+\frac{a^2}{2}\pa{\phi_0'^2-V(\phi_{0})a^2}}\right.\nonumber\\
&+&\left.\sum_{i=1,2}\sum_{k\neq 0}^\infty\paq{v_{i,k}'(\eta)^2+\pa{-k^2+\frac{z''}{z}}v_{i,k}(\eta)^2}\right.\nonumber\\
&+&\left.\frac{1}{2}\sum_{i=1,2}\sum_{\lambda=+,\times}\sum_{k\neq 0}^\infty\paq{\pa{\frac{v_{i,k}^{(\lambda)}}{d\eta}}^2+\pa{ -k^2 +\frac{a''}{a}}\pa{v_{i,k}^{(\lambda)}}^2 }
\right\}
\ea
where $z\equiv \phi_0'/H$ and $H=a'/a^2$ is the Hubble parameter and $L^3\equiv \int d^3x$. The interval $ds$ has dimension of a length $l$ and one generally may either take $[a]=l$ and $[dx]=[d\eta]=l^0$ or $[a]=l^0$ and $[dx]=[d\eta]=l$. Correspondingly one then has $[L]=l^0$ or $[L]=l$. One can eliminate the factor $L^3$ by replacing $a\rightarrow a/L$, $\eta\rightarrow \eta L$, $v\rightarrow \sqrt{L} v$ and $k\rightarrow k/L$. Such a redefinition is equivalent to setting $L=1$ in the above action (\ref{act}) (then implicitly assuming the convention $[a(\eta)]=l$ and $[dx]=[d\eta]=l^0$) and then proceeding with its quantization. Such a choice, although limited to the homogeneous part, has been previously illustrated \cite{FVV}. 
Henceforth we shall use this latter simplifying choice. Only at the end, in order to compare our results with observations we shall restore all quantities to their original definition and the dependence on $L$ will become explicit. 
Let us finally note that the fact that $L$ is infinite does not create a problem. As usual the transition from the Fourier integral w.r.t. the wave number to the Fourier series eliminates the correspondent divergence.  
The dynamics of each d.o.f. describing the perturbations is formally analogous to that of a scalar field with a time dependent mass.  The canonical quantization of the action (\ref{act}) leads to the following Wheeler-De Witt (WDW) equation \cite{DeWitt} for the wave function of the universe (matter plus gravitation)
\begin{eqnarray}
&&\left\{\frac{1}{2\M^2}\frac{\partial^2}{\partial a^2}-\frac{1}{2a^2}\frac{\partial^2}{\partial \phi_0^2}+Va^4\right.\nonumber\\
&&\left.+\sum_{k\neq 0}^{\infty}\paq{-\frac{1}{2}\frac{\partial^2}{\partial v_k^2}+\frac{\omega_k^2}{2}v_k^2}\right\}\Psi\pa{a,\phi_0,\{v_k\}}=0
\end{eqnarray}
where, without lose of generality, just one d.o.f. has been singled out. In particular $\omega_{k}^{2}\equiv k^{2}+m^{2}(\eta)$ and $m^{2}(\eta)=-\frac{z''}{z}$ for each scalar perturbation and $m^{2}(\eta)=-\frac{a''}{a}$ for each tensor perturbation. On performing a Born-Oppenheimer decomposition \cite{BO} for the full WDW equation  one can then obtain the Schwinger-Tomonaga equation for the wave function of each mode of the perturbation $v$ \cite{BO-cosm} on following exactly the step by step derivation described in \cite{K}.
Finally the differential master equation governing the evolution of the two point function
\be{two}
p_k(\eta)\equiv \!\!\!\!\phantom{A}_{s}\langle 0|\hat v_k^{2}|0\rangle_{s}=\av{\hat v_k^{2}}_{0}
\ee
can be derived:
\be{qfx}
\frac{d^{3}p}{d\eta^{3}}+4\omega^{2}\frac{dp}{d\eta}+2\frac{d\omega^{2}}{d\eta}p+\Delta_p=0
\ee
with
\ba{qfxP}
\!\!\!\!\!\Delta_p&\!\!\!\!\!=\!\!\!\!\!&-\frac{1}{\M^{2}}\frac{d^{3}}{d\eta^{3}}\frac{\left(p'^{2}+4\omega^{2}p^{2}-1\right)}{4 a'^{2}}+\frac{1}{\M^{2}}\frac{d^{2}}{d\eta^{2}}\frac{p'\left(p'^{2}+4\omega^{2}p^{2}+1\right)}{4 pa'^{2}}\nonumber\\
&&+\frac{1}{\M^{2}}\frac{d}{d\eta}\left\{\frac{1}{8a'^{2}p^{2}}\left[\left(1-4\omega^{2}p^{2}\right)^{2}+2p'^{2}\left(1+4\omega^{2}p^{2}\right)+p'^{4}\right]\right\}\nonumber\\
&&-\frac{1}{\M^{2}}\frac{\omega\omega'\left(p'^{2}+4\omega^{2}p^{2}-1\right)}{a'^{2}}
\ea
where the subscript $k$ has been omitted.
The above equation is exact to the first order in $\M^{-2}$ and in the $\M\rightarrow \infty$ limit it reproduces the standard evolution of the two point function. Let us note that the above master equation is only valid within a perturbative approach and $\Delta_p$ is the expression for the quantum gravitational corrections to order $\M^{-2}$ to the evolution of the Bunch-Davies (BD) vacuum \cite{BD} (see \cite{K} for more details). 
\section{Slow-Roll Inflation}
The de Sitter evolution is a fairly good approximation to the inflationary dynamics. It has been studied in detail in paper \cite{K} where the equation (\ref{qfx}) was solved exactly for the scalar perturbations. In the standard approach (where the quantum gravitational corrections are neglected) the dynamics of the tensor perturbations is exactly the same as that for the scalar sector in the de Sitter case. However if one wishes to calculate the detailed features of the spectra of cosmological perturbations in order to compare them with observations one must go beyond the de Sitter approximation.  A more refined but still approximate approach to the evaluation of the spectrum of cosmological perturbations is that of slow-roll. Such an approach is not associated with a well defined gravity-inflaton action but describes well the evolution of cosmological perturbations during a generic inflationary phase having a slowly varying Hubble parameter and scalar field. Realistic inflationary models are treated in the slow-roll (SR) approximation and the features of the spectra of perturbations generated during inflation are accurately estimated  in such a framework. It is then worth generalizing our procedure to such a case.\\
The SR approximation is based on the assumption that the inflaton field $\phi_0(t)$ and the Hubble parameter $H(t)$ vary slowly during the inflationary period when cosmological perturbations cross the horizon. Slow rolling can be described on introducing the so-called SR parameters. There exist different families of SR parameters associated, for example, with inflaton evolution, with Hubble parameter evolution or with the inflaton potential. In particular we indicate by $\epsilon_i$ with $i=1,\dots,+\infty$ the hierarchy associated with the evolution of $H(t)$ defined by
\be{ephierarchy}
\epsilon_0\equiv \frac{H_0}{H(t)}\;,\quad \epsilon_{n+1}=\frac{1}{\epsilon_n}\frac{d\epsilon_n}{d\ln a}
\ee
and we indicate by $\delta_i$ the hierarchy associated with the evolution of $\phi_0(t)$ where
\be{dehierarchy}
\de{0}\equiv \frac{\phi(t)}{\phi_0}\;,\quad \de{n+1}=\frac{1}{\de{n}}\frac{d\de{n}}{d\ln a}.
\ee
The SR approximation consists in assuming $|\ep{i}|\ll1$ and $|\de{i}|\ll1$. In such a regime the logarithmic derivative w.r.t. $a$ of each of these parameters is negligible to first order in the parameters themselves. Then on neglecting second order contributions one can treat the parameters as constants and find approximate solutions to the equations governing the dynamics of the cosmological perturbations.
We further note that through the Einstein equations one can generally find various relations among the different hierarchies at least to the first order in the parameters.\\
Only few of the above parameters generally appear in the equations for the perturbations arising from the underlying theory. In the GR framework it is quite common to introduce the SR parameters
\be{SRlit}
\epsilon_{SR}\equiv -\frac{\dot H}{H^2}=\ep{1}\;\;{\rm and}\;\; \eta_{SR}\equiv-\frac{\ddot \phi_0}{H\dot \phi_0}=\ep{1}-\de{2}-\de{1}
\ee
and obtain the spectra just in terms of these two. 
To first order in the SR approximations the scale factor evolution satisfies the equation
\be{aSReq}
a H\simeq-\frac{1+\ep{1}}{\eta}
\ee
where higher orders in ${\epsilon_{i}}$ have been neglected. Its solution is then given by
\be{aSR}
a=a_0\pa{\frac{\eta_0}{\eta}}^{1+\ep{1}}.
\ee
In terms of the above quantities one finds
\be{omegas}
\omega^2=k^{2}-\frac{2\pa{1+3\epsilon_{SR}-\frac{3}{2}\eta_{SR}}}{\eta^{2}}
\ee
for the scalar perturbation and 
\be{omegat}
\omega^2=k^{2}-\frac{2\pa{1+\frac{3}{2}\epsilon_{SR}}}{\eta^{2}}
\ee
for the tensor perturbations. In equation (\ref{qfx}) the SR parameters appear in the expressions for $\omega$ and $a(\eta)$. To the first order in the SR approximation they are small and can be treated as constants. Let us note that, owing to the forms of (\ref{omegas}) and (\ref{omegat}), it is possible to recover the equation for the tensor perturbations starting from the equation for the scalar perturbations and taking the limit $\eta_{SR}\rightarrow \ep{SR}$.
\section{Evolution of the scalar perturbations}
As already mentioned the correct treatment of the quantum gravitational corrections involves the evaluation of $\Delta_p$ using the unperturbed BD solution. Such a solution can be expressed in a conventional way in terms of a product of Hankel functions 
\be{pHank}
p(\eta)=-\frac{\pi \eta}{4}H_\nu^{(1)}(-k\eta)H_\nu^{(2)}(-k\eta)
\ee
with $\nu=\frac{3}{2}+2\ep{SR}-\eta_{SR}$ \cite{Stewart:1993bc}. The expression for $\Delta_p$ turns to be a very complicated function of the above Hankel functions and their derivatives. We therefore evaluate $\Delta_p$ in the short and in the long wavelength limits and simply keep the leading terms for each power of $-k\eta$. This approximation is reasonable considering the smallness of the SR parameters. In such an approximation we neglect irrelevant small deviations from the de Sitter results proportional to the SR parameters and just keep the new peculiar effects originating from the SR dynamics which can be comparable with the de Sitter terms. In the short wavelength limit (where the initial conditions are fixed) the expression for the quantum corrections is given by
\be{corrSW}
\Delta_p^{(S)}\simeq \frac{\pa{\frac{\eta}{\eta_0}}^{2(1+\ep{SR})}}{a_0^2\M^2\eta}\paq{-\frac{2\ep{SR}}{\pa{-k\eta}^2}+\frac{4}{\pa{-k\eta}^4}}
\ee
and in the long wavelength limit one finds
\be{corrLW}
\Delta_p^{(L)}\simeq\frac{\pa{-2k\eta}^{-4\pa{2\ep{SR}-\eta_{SR}}}\pa{\frac{\eta}{\eta_0}}^{2(1+\ep{SR})}}{a_0^2\M^2 k^4\eta^5}\paq{\frac{7\pa{\eta_{SR}-\ep{SR}}}{\pa{-k\eta}^2}+4}.
\ee
The terms in the brackets with coefficients independent of the SR parameters reproduce the correct de Sitter behavior for $\ep{SR},\eta_{SR}\rightarrow 0$
\be{deltaDS}
\Delta_p^{(DS)}=\frac{4H^2}{\M^2 k^4\eta^3}.
\ee
The remaining terms have a pure SR origin and have important consequences in spite of the smallness of the SR parameters. The terms proportional to SR parameters in Eqs. (\ref{corrSW}) and (\ref{corrLW}) dominate over the pure de Sitter contribution (\ref{deltaDS}) in the short and long wavelength limits respectively.
\subsection{Analytical solution}
The simplified expressions for the quantum gravitational corrections (\ref{corrSW},\ref{corrLW}) in the short and in the long wavelength regime are quite simple and accurate with respect to their exact expression in terms of the Hankel functions. The smaller the SR parameters the better is the accuracy of the approximations. One can find the general solutions of Eq. (\ref{qfx}) on solving the equation in these two regimes and then matching the resulting solutions at horizon crossing. 
On changing the independent variable $\eta\rightarrow \z\equiv -\ln\pa{-k\eta}$ equation (\ref{qfx}) takes the form
\ba{qfxSW}
&&\frac{d^3p}{d\z^3}+3 \frac{d^2p}{d\z^2}+\paq{4{\rm e}^{-2\z}-6\pa{1+4\ep{SR}-2\eta_{SR}}}\frac{dp}{d\z}-4\pa{2+6\ep{SR}-3\eta_{SR}}p\nonumber\\
&&+A {\rm e}^{2\ep{SR}\z}\paq{2\ep{SR}{\rm e}^{-2\z}-4}=0
\ea
in the short wavelength limit and
\ba{qfxLW}
&&\frac{d^3p}{d\z^3}+3 \frac{d^2p}{d\z^2}+\paq{4{\rm e}^{-2\z}-6\pa{1+4\ep{SR}-2\eta_{SR}}}\frac{dp}{d\z}-4\pa{2+6\ep{SR}-3\eta_{SR}}p\nonumber\\
&&-\frac{A}{2^{4\pa{2\ep{SR}-\eta_{SR}}}}{\rm e}^{\pa{6\ep{SR}-4\eta_{SR}}\z}\paq{4+7\pa{\eta_{SR}-\ep{SR}}{\rm e}^{2\z}}=0
\ea	
in the long wavelength limit with
\be{Adef}
A=\frac{1}{a_0^2\M^2k^2\pa{-k\eta_0}^{2\pa{1+\ep{SR}}}}.
\ee
Let us emphasize that the expressions (\ref{corrSW}), (\ref{corrLW}) both contain the term which correctly reproduces the de Sitter limit when $\ep{SR}\rightarrow 0$, $\eta_{SR}\rightarrow 0$. Such a term is sub-leading but is necessary to accurately reproduce the short/long wavelength behavior of the quantum gravitational corrections at $-k\eta\simeq 1$.\\
The general solution 	of Eqs. (\ref{qfxSW}), (\ref{qfxLW}) is given by the general solution of the homogeneous equation plus the particular solution of the full equation. In the short wavelength limit an approximate particular solution can be found starting from the ansatz
\be{ansS}
p^{(S)}\simeq A {\rm e}^{-2\ep{SR}\z}\pa{\alpha_S+\beta_S {\rm e}^{2\z}}.
\ee
then obtaining
\be{solpS}
\alpha_S=\frac{1}{4},\;\beta_S=\frac{3}{4}
\ee
to the leading order in SR for each parameter.
In the long wavelength limit we take the following ansatz for the particular solution
\be{ansL}
p^{(L)}\simeq \frac{A}{2^{4\pa{2\ep{SR}-\eta_{SR}}}} {\rm e}^{\pa{6\ep{SR}-4\eta_{SR}}\z}\pa{\alpha_L {\rm e}^{2\z}+\beta_L}.
\ee
and we obtain
\be{solpL}
\alpha_L=-\frac{7}{36}, \;\beta_L=-\frac{25}{36}.
\ee
The approximate particular solutions (\ref{ansS}) and (\ref{ansL}) have a different behavior in the de Sitter limit. Such a difference is a consequence of the existence of an infinite set of particular solutions to inhomogeneous differential equations. The difference between these solutions is equal to a solution of the corresponding homogeneous equation. We can match the short and the long wavelength expressions (\ref{ansS}), (\ref{ansL}) by adding
\be{matchplong}
p_0^{(L)}\simeq A\frac{17}{18}2^{2\pa{2\ep{SR}-\eta_{SR}}}{\rm e}^{2\pa{2\ep{SR}-\eta_{SR}}\z}\pa{1+{\rm e}^{2\z}}
\ee
to (\ref{ansL}). The expression (\ref{matchplong}) is an exact solution of the homogeneous equation in the long wavelength limit. We then replace the solution (\ref{ansL}) by
\ba{solpLM}
p^{(L)}&\simeq& -\frac{A}{18}\left[2^{-4\pa{2\ep{SR}-\eta_{SR}}} {\rm e}^{\pa{6\ep{SR}-4\eta_{SR}}\z}\pa{\frac{7}{2} {\rm e}^{2\z}+\frac{25}{2}}\right.\nonumber\\
&&\left.-17\times2^{2\pa{2\ep{SR}-\eta_{SR}}}{\rm e}^{2\pa{2\ep{SR}-\eta_{SR}}\z}\pa{1+{\rm e}^{2\z}}\right].
\ea
Now (\ref{solpLM}) and (\ref{solpS}) reproduce the same de Sitter solution when $\ep{SR}=\eta_{SR}=0$ and, to zeroth order in SR, the solutions (\ref{solpLM}) and (\ref{solpS}) match at the horizon. To this extent they are two branches of the same particular solution $\Delta p$.\\
Given this particular solution, $\Delta p$, one can finally write the full general solution of (\ref{qfx}) as
\be{fullsol}
p=\frac{\pi }{4k}\pa{1+A\delta}{\rm e}^{-\z}H_\nu^{(1)}({\rm e}^{-\z})H_\nu^{(2)}({\rm e}^{-\z})+\Delta p
\ee
where a residual integration constant $\delta$ has been added. Such a constant multiplied by the factor $A\propto\M^{-2}$ expresses the residual freedom in fixing the initial conditions (vacuum) to order $\M^{-2}$. This also occurs in the pure de Sitter case.
\subsection{Vacuum choice and results}
The vacuum prescription, i.e. the initial condition for the evolution of the modes, is usually fixed in the short wavelength regime when modes are well inside the horizon. The general BD prescription is equivalent to the requirement $p(\z)\simeq 1/\pa{2k}$ when $\z\rightarrow -\infty$. When quantum gravitational corrections are absent in the pure de Sitter case there is a suitable choice of $\delta$ which reproduces the BD initial condition for $\z\rightarrow -\infty$. This statement can be easily verified from our expressions. In the short wavelength limit, on setting the SR parameters to zero, the solution (\ref{fullsol}) takes the form 
\be{dssol}
p^{DS}\sim\pa{1+\frac{H^2}{\M^2 k^4}\delta}\frac{1}{2k}+\frac{H^2}{4\M^2 k^4}.
\ee
The BD vacuum is recovered when $\delta=-k/2$. Consequently the long wavelength limit of the evolved BD vacuum is
\be{BDevoDS}
p^{DS}\sim \paq{1+\frac{H^2}{\M^2 k^4}\pa{-\frac{k}{2}}}\frac{{\rm e}^{2\z}}{2k}+\frac{3H^2}{4\M^2 k^4}{\rm e}^{2\z}=\frac{{\rm e}^{2\z}}{2k}\pa{1+\frac{H^2}{\M^2 k^3}}
\ee
and just corresponds to the results obtained in \cite{K}.\\
When SR dynamics is taken into account an approximate constant solution in the $\z\rightarrow -\infty$ limit no longer exists as the leading part of the particular solution is 
\be{partshort}
p^{(S)}\simeq \frac{A}{4} {\rm e}^{-2\ep{SR}\z}=\frac{A}{4} {\pa{-k\eta}}^{2\ep{SR}}
\ee
and is time dependent.\\
A prescription to fix the vacuum at some time during the evolution of each mode would determine the parameter $\delta$ (in general as a function of $k$). The long wavelength limit of the full solution is
\ba{longwavep}
p^{(L)}&=&\pa{1+\delta A}\frac{1}{8k}\frac{\Gamma(\nu)^2}{\Gamma\pa{3/2}^2}\pa{\frac{-k\eta}{2}}^{-2(1+2\ep{SR}-\eta_{SR})}-\frac{17}{18}A\left[\frac{7}{34}2^{-4\pa{2\ep{SR}-\eta_{SR}}}\right.\nonumber\\
&&\left.\times  {\pa{-k\eta}}^{-2\pa{1+3\ep{SR}-2\eta_{SR}}}-2^{2\pa{2\ep{SR}-\eta_{SR}}}{\pa{-k\eta}}^{-2\pa{1+2\ep{SR}-\eta_{SR}}}\right].
\ea
Just keeping the first order SR coefficients in the exponents of $-k\eta$ in (\ref{longwavep}) is sufficient to analyze the interesting features of the full solution $p$. It can be rewritten as
\be{papprox}
p^{(L)}\simeq \frac{1}{2k}\pa{-k\eta}^{-2\pa{1+2\ep{SR}-\eta_{SR}}}\paq{1+Ak\pa{\frac{\delta}{k}+\frac{17}{9}-\frac{7}{18}\pa{-k\eta}^{-2\pa{\ep{SR}-\eta_{SR}}}}}.
\ee
The coefficient $A$ is approximately
\be{Aapp}
A\simeq \frac{H^2}{\M^2k^{4}}
\ee
where $H$ is the Hubble constant evaluated at some time during inflation (small deviations due to SR are neglected). 
Let us note that $p$ is the two point function associated with the perturbation $v$ which includes a factor $1/\sqrt{L}$ and is expressed in terms of the corresponding set of rescaled quantities. If one now returns to the original physical quantities by substituting $v\rightarrow v/\sqrt{L}$ (and thus $p\rightarrow p/L$), $k\rightarrow L k$, $a\rightarrow L a$, $\eta\rightarrow \eta/L$ one has that $k\eta$ remains invariant and an overall factor $L^{-1}$ appears on both sides of Eq. (\ref{papprox}) and cancels. The quantum gravitation contribution which is mainly encoded in the $A \cdot k$ term changes since $A \cdot k\rightarrow A\cdot k/L^3$. This is the only replacement one needs in order to return to the original, physical, degrees of freedom.  The scale $L\equiv \bar k^{-1}$ would then appear in the result as an effect of the initial integration of the homogeneous dynamics. Analogous results are obtained for the tensor case. We shall return to this later.

Thus, for $k$ large, the quantum gravitational corrections are negligible. For $k$ small, small (perhaps observable) deviations  from the standard SR predictions arise. Depending on $-k\eta$ and the exponent $-2\pa{\ep{SR}-\eta_{SR}}$ such deviations would lead to a power loss in the spectrum of scalar perturbations independently of $\delta$. Let us note that such a possible power loss is a particular consequence of the SR dynamics. In the pure de Sitter the power loss can only originate form some vacuum prescription other than pure BD.\\
On setting $\delta=-k/2$, as in the de Sitter case, one finds 
\be{papproxd1}
p\simeq \frac{1}{2k}\pa{-k\eta}^{-2\pa{1+2\ep{SR}-\eta_{SR}}}\paq{1+\frac{Ak}{18}\pa{25-7\pa{-k\eta}^{-2\pa{\ep{SR}-\eta_{SR}}}}}.
\ee
leading to a power loss if $\ep{SR}>\eta_{SR}$ and for $-k\eta$ small enough. We further note that if $\delta\le-\frac{17}{9}k$ the quantum gravitational corrections always lead to a power loss in the spectrum. The latter prescription leads to a power loss in the de Sitter limit as well and does not depend on SR but is related to the initial conditions of the evolution of the perturbations.

\section{Tensor perturbations}
In the de Sitter and the Power-Law cases, the evolution of the tensor perturbations is governed by the same equation as determines the dynamics of the scalar sector (at least in GR). Differences appear for the case of slow roll. In such a case the effective mass of the scalar perturbation is given by
\be{ddzoverzSR}
\frac{z''}{z}\simeq\frac{2}{\eta^2}\pa{1+3\ep{SR}-\frac{3}{2}\eta_{SR}}
\ee
while for tensor perturbation one has
\be{ddaoveraSR}
\frac{a''}{a}\simeq\frac{2}{\eta^2}\pa{1+\frac{3}{2}\ep{SR}}.
\ee
The unperturbed solution for $p$ is given by (\ref{pHank}) with
\be{nuSRten}
\nu=\frac{3}{2}+\ep{SR}.
\ee
Let us note that the results obtained for scalar perturbations in the SR approximation can be easily changed into the corresponding results for tensor perturbations on letting $\eta_{SR}\rightarrow \ep{SR}$. In such a limit the peculiar behavior found in the scalar sector disappears. The particular solution associated with these corrections is 
\be{solpLMT}
p^{(L)}\simeq -\frac{A}{18}{\rm e}^{2\ep{SR}\z}\left[2^{-4\ep{SR}}\pa{\frac{7}{2} {\rm e}^{2\z}+\frac{25}{2}}-17\times2^{2\ep{SR}}\pa{1+{\rm e}^{2\z}}\right]
\ee
in the long wavelength regime. 
The general solution of (\ref{qfx}) is finally given by the approximate expression
\be{papproxT}
p\simeq \frac{1}{2k}\pa{-k\eta}^{-2\pa{1+\ep{SR}}}\paq{1+Ak\pa{\frac{\delta}{k}+\frac{3}{2}}}
\ee
for $-k\eta\rightarrow 0$ and on neglecting SR parameter dependence as in expression (\ref{papprox}).
On setting $\delta=-1/(2k)$ (BD prescription) one finds
\be{papproxTd1}
p\simeq \frac{1}{2k}\pa{-k\eta}^{-2\pa{1+\ep{SR}}}\paq{1+Ak}
\ee
which leads to a power increase in the small $k$ region. Primordial gravitational waves have not been detected till now. Such an increase, however, would suggest that some signal may be observed in the lowest multipoles interval first if the BD assumption is correct. 

\section{Comparison with observations}
\begin{figure}[t!]
\centering
\epsfig{file=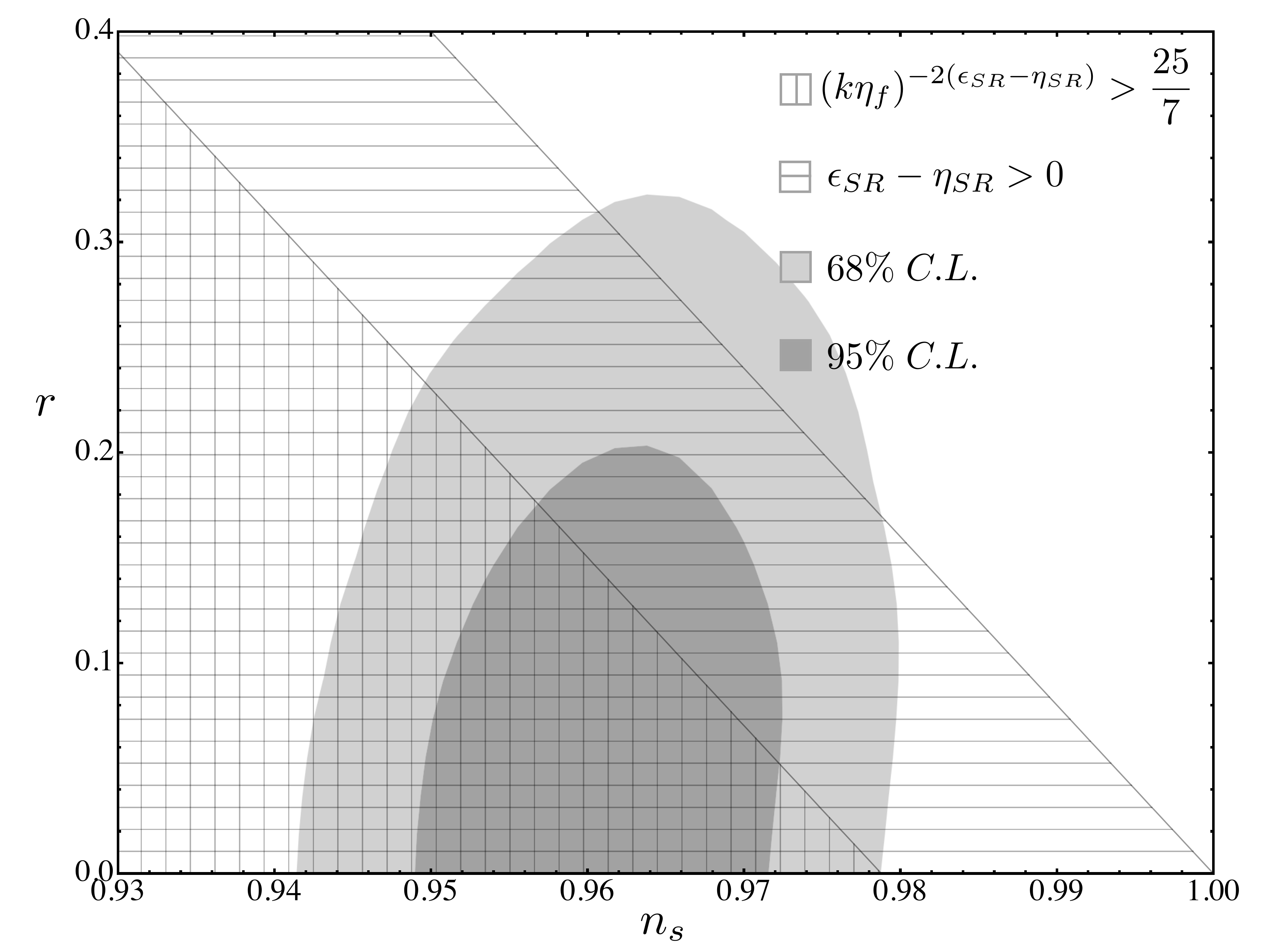, width=11 cm}
\caption{\it Marginalized joint 68\% (darker) and 95\% CL regions for $(r,n_s)$, using Planck+WP+BAO with a running spectral index. The region shaded with horizontal lines is that for which $\ep{SR}-\eta_{SR}>0$, the region shaded with vertical lines is where $(-k\eta_f)^{-2(\epsilon_{SR}-\eta_{SR})}>\frac{25}{7}$.}
\label{fig1}
\end{figure}

\begin{figure}[t!]
\centering
\epsfig{file=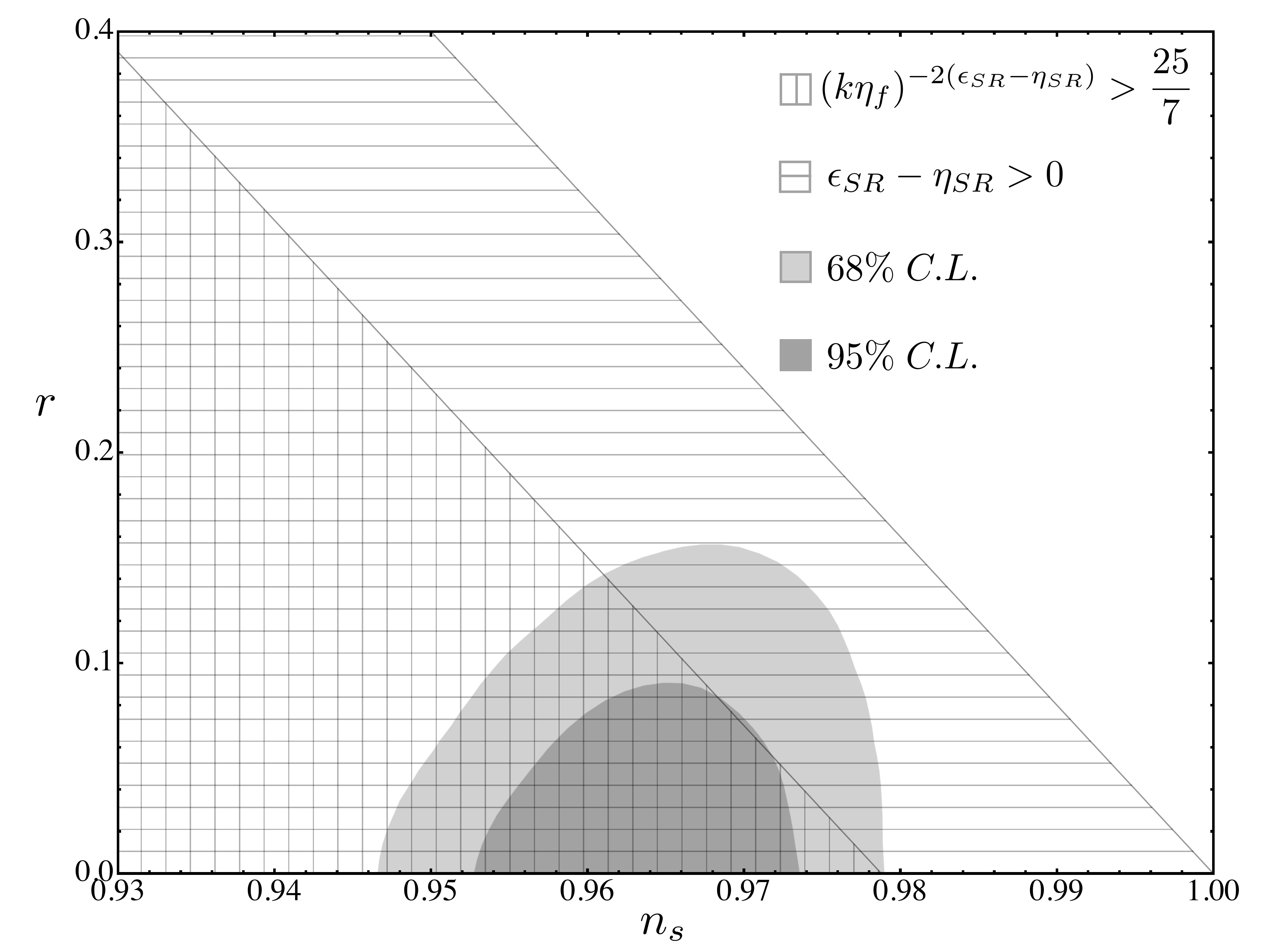, width=11 cm}
\caption{\it Marginalized joint 68\% (darker) and 95\% CL regions for $(r,n_s)$, using Planck+WP+BAO without a running spectral index. The region shaded with horizontal lines is that for which $\ep{SR}-\eta_{SR}>0$, the region shaded with vertical lines is where $(-k\eta_f)^{-2(\epsilon_{SR}-\eta_{SR})}>\frac{25}{7}$.}
\label{fig2}
\end{figure}

The scalar perturbations are usually described in terms of the quantity
\be{Rdef}
\mathcal{R}_k\equiv \frac{H}{\dot \phi_0}\delta\phi= -\frac{v_k}{a}\frac{H}{\dot \phi_0}
\ee
namely the comoving curvature perturbation $\mathcal R$ in the uniform curvature gauge. In the long wavelength limit the power-spectrum associated with $\mathcal R$ is 
\be{PR}
\mathcal{P}_{\mathcal{R}}=\frac{k^3}{2\pi^2}|\mathcal{R}_k|^2=\frac{k^3}{2\pi^2}\frac{H^2}{\phi_0'^2}p_s^{(L)}
\ee
where $p_s^{(L)}$ is given by (\ref{papprox}) and the subscript $s$ indicates the ``scalar sector''.
When the quantum gravitational corrections are neglected, the above spectrum (\ref{PR}) is constant to first order in SR. When quantum corrections are not negligible a time dependence due to the contribution proportional to the $(-k\eta)^{-2\pa{\ep{SR}-\eta_{SR}}}$ term in the brackets of expression (\ref{papprox}) is present. The overall amplitude of the scalar perturbations can be evaluated at the end of the inflationary era when $\eta=\eta_f$. The modes we observe today left the horizon during the inflationary era around 60 e-folds before inflation ended. For these modes $-k\eta_f\sim {\rm e}^{-60}$ up to small corrections proportional to the SR parameters. In the figures (\ref{fig1}), (\ref{fig2}) we show that a power loss for small $k$ with the BD vacuum prescription leading to expression (\ref{papproxd1}) is obtained with a set of SR parameters compatible with observations. We also assume that the evolution of the spectrum of the scalar perturbations after inflation ends does not significantly alter this power loss, at least qualitatively. 

The spectrum of the tensor perturbations 
\be{detens}
|h_k^{(\lambda)}|=\sqrt{2}/(a\M)\sqrt{\sum_{i=1,2}\pa{v_{i,k}^{(\lambda)}}^2}
\ee
is defined as
\be{Ph}
\mathcal{P}_h=\frac{k^3}{2\pi^2}\pa{|h_k^{(+)}|^2+|h_k^{(\times)}|^2}=\frac{4k^3}{\pi^2a^2\M^2}p_t^{(L)},
\ee
where $p_t^{(L)}$ is given by (\ref{papproxT}) and the subscript $t$ indicates the ``tensor sector''. Let us note that the spectrum (\ref{Ph}) is still constant on including the quantum gravitational corrections. 
The above amplitudes (\ref{PR}), (\ref{Ph}) define the tensor to scalar ratio which is usually evaluated in the leading order in the SR approximation and is given by
\be{rdef}
r=\frac{\mathcal{P}_h}{\mathcal{P}_{\mathcal R}}=16|\ep{SR}|
\ee
when quantum gravitational corrections are neglected. The tensor to scalar ratio is modified by quantum gravitational corrections and in particular, in the low $k$ region, a power enhancement in the numerator may be either balanced by a power enhancement or counterbalanced by a power loss in the denominator. In the latter case an overall increase in the ratio for $k$ small occurs.
On neglecting the quantum corrections one finds the standard results for the spectral indices and in particular
\be{nsm1}
n_s-1=2\eta_{SR}-4\ep{SR}
\ee
and
\be{nt}
n_t=-2\ep{SR}.
\ee
During inflation one generally has $\dot H<0$ and consequently $\ep{SR}>0$. One then has the following relation
\be{conrel}
r=-8n_t=16\ep{SR}.
\ee
The values of $\ep{SR}$ and $\eta_{SR}$ depend on $(n_s,r)$ through equations (\ref{nsm1}) and (\ref{conrel}). In particular we are interested in the sign of $\ep{SR}-\eta_{SR}$ which appears throughout our results. The area shaded with horizontal lines in figures (\ref{fig1}), (\ref{fig2}) plots the region where $\ep{SR}-\eta_{SR}>0$ and compares with the observed values for $n_s$ and $r$. 
Let us note that the quantum gravitational corrections in the expressions (\ref{papprox}) and (\ref{papproxT}) have the form of a few additive, $k$-dependent terms. The wavenumber $k$ is dimensionless in our notation. In GR $k$ may have the dimensions of an inverse length depending on the conventions used. The final results must be consistent with both the conventions. 
Let us remember that in this paper the scale factor, which is quantized, has the dimensions of a length thus leading to a dimensionless $k$ and we have already illustrated in section 2 the reasons for such a choice.
In order to obtain the final results in terms of the original physical quantities we must again reintroduce the scale $\bar k$ ($L^{-1}$) as we explicitly illustrated at the end of section 4. This reintroduction then makes our results independent on the conventions used. 
The value of $\bar k^{-1}$ is a length scale and is usually taken to be the largest observable scale of the universe today \cite{Calcagni}. If so, it is then comparable with the pivot scale $k_*$ which in the case of the Planck data analysis has been taken to be $k_*\simeq 0.05\;{\rm Mpc}^{-1}$ \cite{cmb}.\\
The quantum gravitational corrections determine a deviation with respect to the standard results (without quantum gravitational corrections) proportional to the time independent coefficient $A\cdot k$ which can be rewritten, after some algebra, as
\be{Akest}
A\cdot k=\pa{\frac{\bar k}{k}}^3\frac{H^2}{\M^2\pa{1+\ep{SR}}^2\pa{-k\eta}^{2\ep{SR}}}.
\ee
Let us note that the ratio $\bar k/k$ is unchanged if physical wavenumbers $\bar k/a$ and $k/a$ are used and, of course, $k\eta\simeq k/(aH)$, is independent of the applied conventions.
The time dependence of the numerator and that of the denominator exactly cancel to the first order in SR. Then (\ref{Akest}) can be evaluated at any time during inflation and in particular at the horizon crossing when $-k\eta\simeq 1$ and $H\simeq H_*$. To the leading order in SR we then obtain
\be{Akest2}
A\cdot k\simeq\pa{\frac{\bar k}{k}}^3\frac{H_*^2}{\M^2}.
\ee
The ratio $\frac{H_*^2}{\M^2}\le 10^{-6}$ is related to the amplitude of the scalar perturbations and is very small. Thus $\left.A\cdot k\right|_{k\sim k_*}$ is tiny if we take $\bar k\simeq 1.4\cdot 10^{-4}\,{\rm Mpc}^{-1}$ to be the largest observable scale in the universe today. On the other hand if $\bar k\sim k_*$ the amplitude of the quantum gravitational corrections is bigger but still too small to be observed given the precision of present day experiments. For the case of tensor perturbations the results are similar and the same considerations hold.\\
Let us however note that the results obtained so far are based on the assumption that SR approximation is valid during the evolution of each observable mode. What happens between the last stages of inflation (when SR no longer applies) and today is still unspecified as far as the quantum gravitational part of the spectrum is concerned. A tiny power loss or enhancement may well be dampened or enhanced during the evolution till now and its effect may then be observed in Planck data. The evolution of such corrections from the end of inflation until today is an ambitious task which goes beyond the scope of this paper. 
\section{Conclusions}
In this paper we solved the general master equation ({\ref{qfx}}) describing the lowest order corrections coming from quantum gravitational contributions to the spectrum of cosmological fluctuations on assuming an inflationary evolution generically described by SR dynamics. This letter is a generalization of the previous article \cite{K} where such an equation was obtained through a Born-Oppenheimer decomposition of the inflaton-gravity system and solved exactly for the two point function of the scalar fluctuation for the case of a de Sitter evolution. The more realistic case of an inflationary SR dynamics has been addressed here. The quantum gravitational corrections for the SR case have peculiar features and are very different from the de Sitter case. In particular, for the case of the scalar fluctuations, their form is not simply a deformation of the de Sitter result proportional to the SR parameters. New contributions arise due to SR and their effect dominates over the de Sitter-like contributions for very small and very large wavelengths. The small wavelength region is that which affects the initial state (vacuum) of each mode of the perturbations. The long wavelength region is that associated with the observations of the spectrum of perturbations. The new contributions are proportional to $\ep{SR}-\eta_{SR}$ and are zero for the de Sitter and power-law cases. They can lead to a power-loss term for low $k$ in the spectrum of the scalar curvature perturbations at the end of inflation providing the difference $\ep{SR}-\eta_{SR}>0$. Furthermore the evolution of the primordial gravitational waves has also been addressed. The quantum gravitational corrections also affect the dynamics of tensor perturbations and determine a deviation from the standard results in the low multipole region. 

Finally our analytical results are compared with observations. The quantum gravitational corrections originate a power loss in the scalar spectrum compatible with the Planck constraint on $\ep{SR}-\eta_{SR}$. Another possible source of the power loss is related to the perturbed vacuum choice. An accurate analysis of the possible outcome of some non-standard choices of the vacuum is beyond the scope of this paper and is not addressed here. The amplitude of the quantum gravitational effects depends on the product $A\cdot k$. Unfortunately within the present approach an estimate of this amplitude during inflation leads to a tiny result. Such an estimates has been performed in a conservative manner \cite{Calcagni} by introducing a length scale $\bar k^{-1}$ associated to the size of the observable universe today. Should such corrections freeze at the end of inflation they are probably invisible to present day experiments. Different choices of $\bar k$ should however lead to very different estimates.
Of course the choice of a smaller length scale ($\bar k^{-1}$) will lead to stronger quantum gravitational effects.\\
\\
\section*{Acknowledgments}
The work of A. K. was partially supported by the RFBR grant 14-02-00894. We wish to thank F. Finelli for useful comments.


\end{document}